\documentclass[12pt,a4paper]{article}

\usepackage{amsmath,amssymb}
\usepackage{booktabs}
\usepackage{array}
\usepackage{geometry}
\usepackage{setspace}
\usepackage[hang,small,bf]{caption}
\usepackage{threeparttable}
\usepackage{makecell}
\usepackage{hyperref}
\usepackage{natbib,multirow}

\newcolumntype{R}[1]{>{\raggedleft\arraybackslash}p{#1}}
\newcolumntype{C}[1]{>{\centering\arraybackslash}p{#1}}

\title{\vspace{-2cm}\textbf{The Evolution of Intersectoral Financial Relations between Industry and Agriculture in China's Planned Economy Period\\——Evidence from Fiscal, Price, and Trade Channels}}

\author{Jiyuan Lyu\\
	\small School of Economics and Management, Beijing University of Technology, email:sergeylyu@163.com}

\begin{document}
	\maketitle
	
	\begin{abstract}
		\setstretch{1.0}
		
		This paper examines intersectoral financial relations between industry and agriculture during 1949--1977 through four channels: fiscal, price, quantity procurement, and foreign trade. We find that these relations underwent staged evolution rather than constituting a persistent transfer mechanism: agriculture supported industry in the 1950s, the year 1961 marked a turning point, and the 1970s witnessed the emergence of mutual interdependence between the two sectors. No stable empirical pattern of price mechanisms persistently affecting industrial profitability is observed.
		
		\vspace{0.3cm}
		
		
		\vspace{0.2cm}
		
		
	\end{abstract}
	
	\newpage
	
	\section{Introduction}
	
	The intersectoral financial relationship between industry and agriculture during the planned economy period is one of the most important resource allocation issues in understanding China's industrialization path. Under conditions of limited capital accumulation capacity and an underdeveloped financial system, what resource-supply function the agricultural sector performed in the early stage of industrialization, and through which institutional arrangements the state allocated resources between the two sectors, bear not only on evaluations of the pre-reform industrialization model but also on our understanding of the continuity between the pre- and post-reform eras. Whether a persistent resource transfer existed between industry and agriculture, through what channels such transfer was realized, and whether it remained stable throughout the planned economy period have thus long been central questions in Chinese economic history.
	
	On this question, the ``price scissors'' hypothesis has constituted the most influential explanatory framework. Its theoretical origin can be traced to Preobrazhensky's (1926) discussion of socialist primitive accumulation, namely, that unequal exchange between industrial and agricultural products transfers agricultural surplus to the industrial sector, providing funding for industrialization \citep{preob1965}. This line of reasoning was subsequently widely applied in Chinese economic history research. Yan et al.\ (1990), using the industrial--agricultural price ratio as a basis, systematically estimated the price scissors for 1952--1978 and argued that agriculture had long served as an important source of industrial accumulation \citep{yan1990}; Sheng (1993) further incorporated fiscal, price, and foreign trade channels into a unified analytical framework and similarly concluded that agriculture had continuously transferred resources to the industrial sector \citep{sheng1993}. Although differing in emphasis, Lardy (1983) and Riskin (1987) both held that the agricultural sector bore a relatively heavy resource-supply burden during the planned economy period \citep{lardy1983,riskin1987}.
	
	At the same time, there have long been divergent views regarding the scale of agricultural resource transfers and the mechanisms through which they were realized. Bramall (2008) pointed out that estimates of the price scissors are highly sensitive to the choice of base-period prices, and different reference systems can yield markedly different transfer magnitudes \citep{bramall2008}. Sicular (1988) emphasized that under the system of unified procurement and sale, price formation was jointly influenced by multiple institutional factors, including administrative allocation, fiscal arrangements, and commercial circulation, so that price differentials per se cannot be directly equated with resource flows \citep{sicular1988}. Rawski (1980) further argued that as the industrial sector's self-accumulation capacity gradually strengthened, the sources of industrialization funding may already have changed, and the experience of the 1950s cannot simply be extrapolated to the entire planned economy period \citep{rawski1980}. These discussions suggest that intersectoral financial relations between industry and agriculture may not have been an unchanging institutional arrangement throughout 1949--1977, but rather exhibited evolutionary features in response to economic development and policy adjustments.
	
	Existing research has laid an important foundation for understanding industry--agriculture relations during the planned economy period, but two issues in empirical identification merit further discussion. First, most existing studies estimate resource transfers based on price differentials or sectoral income--expenditure accounts, yet rarely test directly whether different channels actually produced observable capital accumulation effects in the industrial sector. There is no one-to-one correspondence among price distortions, fiscal flows, commercial profits, and industrial profitability; accounting results for one channel do not automatically vindicate the entire mechanism. Second, most studies tend to treat 1949--1977 as a relatively unified analytical period, while discussion of whether institutional environments and financial flows changed across different historical stages remains relatively limited. If intersectoral financial relations are themselves characterized by staged features, then summarizing their operation by period averages risks obscuring important historical differences.
	
	Based on the above considerations, the question this paper addresses is not simply whether the ``price scissors'' existed, but rather how intersectoral financial relations between industry and agriculture evolved during the planned economy period, and through which channels this evolution principally manifested. To this end, we draw on officially published statistical materials for 1949--1977 to systematically examine intersectoral financial flows from fiscal, price, grain procurement, and foreign trade perspectives. In the price channel analysis, we use the capital profit rate of industrial enterprises as an indicator of industrial sector profitability and employ the Toda--Yamamoto causality test to address the problem of uncertain orders of integration in the time series; for the potential endogeneity of raw material dependence, we further apply instrumental variable methods. None of these analyses aims to reach a judgment based on any single indicator; rather, we seek to reconstruct the historical evolution of intersectoral financial relations through mutual corroboration across different channels.
	
	It should be noted that this paper discusses intersectoral financial relations, not an overall assessment of planned economy performance. The contribution of agriculture to industrialization in the 1950s is amply supported by historical evidence, and we have no intention of denying this historical fact. The question at a different empirical level that this paper addresses is: can this contribution be characterized as a persistent resource transfer mechanism spanning the entire planned economy period, or, with changes in economic structure, fiscal institutions, and trade patterns, had intersectoral financial relations already undergone staged adjustments?
	
	The remainder of the paper is organized as follows. Section~2 introduces data sources, variable construction, and research design. Section~3 first employs multiple statistical series to demonstrate the structural changes in intersectoral financial relations around 1961. Sections~4 through~6 analyze the concrete manifestations of this change through the fiscal channel, the price mechanism, and the quantity and trade channels, respectively. Section~7 discusses, against the historical background, the possible mechanisms behind the evolution of intersectoral financial relations. Section~8 summarizes the main findings and discusses the limitations of the paper and their implications for understanding industry--agriculture relations during the planned economy period.
	
	\section{Data and Research Design}
	
	\subsection{Data Sources}
	
	The data used in this paper are drawn primarily from two officially published compilations by the National Bureau of Statistics: \textit{A Comprehensive Compendium of China's Rural Economic Statistics (1949--1986)} and \textit{Statistical Summary of Thirty Years of National Economic Development (1949--1978)}. The study sample covers 1952--1977, with some fiscal and trade data traceable to 1950. Since financial statistics for state-owned independent accounting industrial enterprises have been published continuously since 1952, all econometric analyses involving industrial enterprise profitability use annual data for 1952--1977 ($n=26$).
	
	The data used in this paper fall into five categories. The first comprises fiscal data, including fiscal revenue, fiscal expenditure, and profit and tax indicators by economic sector, used to examine financial flows between agriculture and the state budget. The second comprises industrial enterprise financial data, including net fixed assets, working capital, profits, taxes, gross industrial output value, and sales revenue, used to measure industrial sector profitability. The third comprises agricultural data, including grain output, grain procurement volumes, and major agricultural product price indices, used to analyze both price and quantity channels. The fourth comprises macroeconomic indicators such as national income, consumption, and accumulation, used to examine the links between intersectoral financial relations and overall accumulation. The fifth comprises import and export commodity structure data, including export composition, import composition, and quantities of major traded commodities, used to analyze changes in the foreign trade channel.
	
	The instrumental variables used in the price transmission model---plastic output, chemical fiber supply, and synthetic rubber output---are taken from the industrial product statistics in \textit{Statistical Summary of Thirty Years of National Economic Development}.
	
	\subsection{Core Variables}
	
	The price channel analysis employs the capital profit rate (CPR) of industrial enterprises to measure industrial sector profitability, defined as
	
	\begin{equation}
		\text{CPR}_t=
		\frac{\text{Profit}_t+\text{Tax}_t}
		{\text{Net fixed assets}_t+\text{Working capital}_t}.
	\end{equation}
	
	This indicator is derived directly from enterprise financial statistics and provides a relatively stable measure of the return on industrial enterprise capital. To test whether results are sensitive to the indicator's construction, we also employ the pure capital profit rate (profit/total capital) and the output-value profit rate (profit/gross industrial output value) for robustness comparisons.
	
	The industrial--agricultural terms of trade (TOT) are defined as
	
	\begin{equation}
		\text{TOT}_t=
		\frac{\text{Agricultural and sideline product procurement price index}_t}
		{\text{Rural industrial goods retail price index}_t}
		\times100,
	\end{equation}
	
	where both price indices use 1950 as the base year. An increase in TOT indicates an improvement in the relative price of agricultural products; a decrease indicates deterioration.
	
	The quantity channel is represented by the net grain procurement rate:
	
	\begin{equation}
		q_t=
		\frac{\text{Net grain procurement}_t}
		{\text{Total grain output}_t},
	\end{equation}
	
	where net procurement has already deducted quantities resold to rural areas. To capture changes in both price and quantity dimensions simultaneously, we further construct a composite indicator:
	
	\begin{equation}
		E_t=
		\frac{q_t}{\text{TOT}_t/100},
	\end{equation}
	
	This indicator is primarily used to describe long-run changes in intersectoral resource flows and is not directly employed in subsequent causal analysis.
	
	\subsection{Research Design}
	
	The question this paper addresses is how intersectoral financial relations between industry and agriculture evolved during the planned economy period, rather than merely whether a particular resource transfer channel existed. Accordingly, the paper does not rely on any single indicator, but examines evidence from four aspects---fiscal, price, quantity, and foreign trade---and compares whether their changes over time display consistency.
	
	The fiscal channel primarily reflects financial flows between agriculture and the state budget, including the agricultural sector's fiscal contribution, fiscal expenditure in support of agriculture, and the fiscal balance of the grain procurement and distribution system. If agriculture had long served as a source of industrial accumulation, fiscal flows should remain relatively stable; conversely, if agriculture gradually became a net recipient of fiscal resources, the fiscal revenue and expenditure structure should exhibit corresponding changes.
	
	The price channel focuses on whether relative industrial--agricultural prices translate into observable changes in industrial sector profitability. If the price mechanism constitutes an important mode of intersectoral resource transfer, changes in the terms of trade between industry and agriculture should be reflected in the capital profit rate of industrial enterprises; if a stable link is absent, price differentials alone are insufficient to demonstrate that resource transfer has been realized.
	
	The quantity channel mainly examines changes in the grain procurement system. Grain was the most important commodity under the unified procurement and sale system during the planned economy period, and changes in its procurement ratio can reflect the intensity with which the state regulated industry--agriculture relations through physical allocation.
	
	The foreign trade channel focuses on changes in export and import structures. If agriculture had continuously performed a foreign-exchange-earning function for industrialization, export and import structures should remain relatively stable; conversely, if the industrial sector gradually developed autonomous foreign-exchange-earning capacity, trade structure should adjust accordingly.
	
	These four channels reflect different aspects of intersectoral financial relations. They are mutually independent yet complementary.
	
	\subsection{Econometric Methods}
	
	This paper comprehensively employs three types of methods: structural break tests, time-series causality analysis, and instrumental variable estimation.
	
	First, for major time series such as fiscal, price, and trade indicators, the Quandt--Andrews unknown breakpoint test is applied to identify years in which structural changes may have occurred, in order to examine whether intersectoral financial relations experienced staged adjustments.
	
	Second, ADF, Phillips--Perron (PP), KPSS, and Zivot--Andrews unit root tests are implemented for each variable. Because some variables yield divergent judgments regarding their order of integration, we adopt the Toda--Yamamoto (1995) Granger causality test as the primary causal identification method. This method conducts Wald tests directly in a VAR($p+d_{\max}$) system in levels, requiring neither identical orders of integration across variables nor prior cointegration testing, and is thus suitable for the sample size and data characteristics of this paper. Finite-sample inference employs the wild bootstrap (Rademacher weights, 5,000 replications) to compute $p$-values.
	
	Finally, for the potential endogeneity of raw material dependence in the price transmission model, we apply the Hausman test for diagnosis and use plastic output, chemical fiber supply, and synthetic rubber output as instrumental variables in two-stage least squares (2SLS) estimation, to test the sensitivity of the main conclusions to potential endogeneity.
	
	All data processing and econometric analysis were carried out in Python, primarily using \texttt{statsmodels} and author-written bootstrap programs.
	
	\section{Staged Changes in Intersectoral Financial Relations}
	
	Before discussing the specific mechanisms of intersectoral financial relations, it is necessary to first address a more fundamental question: Can 1949--1977 be regarded as a unitary period with stable institutions and financial flows? If the answer is negative, we need to identify when changes emerged and in which dimensions they principally manifested.
	
	Existing research has mostly treated the entire planned economy period as a unified object of analysis, estimating cumulative resource transfers. This approach facilitates estimation of long-run resource flows, but may also obscure institutional differences across historical stages. This paper first compares several key indicators---fiscal, price, grain procurement, and foreign trade---to examine whether intersectoral financial relations display clear staged changes.
	
	From a fiscal perspective, the ratio of state fiscal revenue to national income generally trended upward in the 1950s and reached its highest level of the study period in 1960; thereafter it fell rapidly and stabilized at a lower level over the longer term. At the same time, the agricultural sector's contribution to fiscal revenue declined continuously, while fiscal expenditure in support of agriculture gradually increased, signaling a marked shift in agriculture's position within the fiscal system.
	
	The price system exhibited similar temporal features. In 1961, the state substantially raised the procurement prices of major agricultural products, and the industrial--agricultural terms of trade (TOT) registered their largest single-year improvement of the study period. Thereafter, the relative price of agricultural products remained generally above the levels of the 1950s and did not revert to the earlier price relationship. This adjustment in price policy forms a fairly distinct dividing line with the 1950s.
	
	The grain procurement and distribution system also underwent important changes. The grain procurement rate reached an abnormally high level in 1959--1960, then gradually declined after 1961; the operating position of state-owned grain enterprises shifted from small profits to persistent losses. This implies that the role of the grain procurement and distribution system within public finance was no longer the same as in the early 1950s.
	
	Changes in foreign trade structure further reflect this adjustment. In the 1950s, China's exports were still dominated by agricultural products and their primary processed goods, while imports were concentrated in machinery and equipment. Around 1961, the share of machinery and equipment in imports dropped markedly, and grain trade shifted from net exports to net imports; entering the 1970s, the importance of petroleum exports and industrial goods exports gradually increased, while the share of agricultural producer goods in the import structure also rose, giving the trade cycle new structural features.
	
	These changes do not derive from a single statistical source but rather from fiscal statistics, industrial enterprise financial statistics, grain procurement and distribution statistics, and customs trade statistics respectively. There is no direct statistical linkage among these different data sources, so their synchronous change over time carries certain empirical significance. To further investigate this phenomenon, we apply the Quandt--Andrews unknown breakpoint test to the major time series.
	
	Table~\ref{tab:breakpoint} summarizes the structural change test results for the main variables.
	
	\begin{table}[htbp]
		\centering
		\caption{Structural Change Tests for Main Variables}
		\label{tab:breakpoint}
		\begin{threeparttable}
			\begin{tabular}{lcc}
				\toprule
				Variable & Identified Breakpoint & Description \\
				\midrule
				Fiscal revenue/national income & 1962 & Fiscal concentration declined after the Great Leap Forward \\
				Machinery \& equipment import share & 1961 & Import structure underwent marked adjustment \\
				Industrial--agricultural TOT & Around 1961 & Agricultural procurement price adjustment \\
				Grain enterprise profits & Around 1961 & Shifted from profits to persistent losses \\
				Net agricultural fiscal flow & Around 1962 & Direction of fiscal revenue--expenditure changed \\
				\bottomrule
			\end{tabular}
			\begin{tablenotes}
				\small
				\item Note: Breakpoints for fiscal revenue/national income and machinery \& equipment import share are from the Quandt--Andrews test; for the remaining variables, due to clear institutional adjustments or relatively short data series, judgments are based primarily on a combination of policy background and statistical series changes, with detailed results provided in the relevant sections below.
			\end{tablenotes}
		\end{threeparttable}
	\end{table}
	
	It should be stressed that these statistical results do not themselves explain why intersectoral financial relations changed. The period around 1961 experienced not only post-Great Leap Forward economic adjustment, but also the impact of changing Sino-Soviet relations, the grain crisis, and fiscal policy adjustments; the timing of breakpoints alone is insufficient to identify specific mechanisms. The purpose of this section is therefore merely to point out that, whether observed from fiscal, price, or foreign trade perspectives, the period around 1961 displays statistical characteristics clearly distinct from those of the 1950s. This implies that treating 1949--1977 as a unitary whole with a unified financial operating mechanism is, at minimum, worthy of further empirical discussion.
	
	The following sections will examine the fiscal, price, quantity, and trade channels separately, analyzing whether the changes described above reflect adjustments in intersectoral financial relations and what roles different mechanisms may have played in this process.
	
	\section{The Fiscal Channel: Changes in Intersectoral Financial Flows}
	
	Public finance was an important channel of resource allocation during the planned economy period, and the most direct window through which to observe intersectoral financial relations. Whether the agricultural sector continuously provided funds to the industrial sector would first be reflected in the structure of fiscal revenue and expenditure. Accordingly, this section does not directly address the price mechanism, but first examines whether financial flows between agriculture and the state budget remained stable.
	
	It should be noted that fiscal data reflect resource flows within the budgetary system and cannot capture extra-budgetary mechanisms such as price distortions. Hence, the focus of this chapter is the changing position of agriculture within the fiscal system, not the entirety of intersectoral financial relations. The price channel will be discussed separately in the next chapter.
	
	\subsection{Changes in Fiscal Revenue}
	
	The ratio of fiscal revenue to national income reflects the overall degree to which the state concentrated resources from the national economy, and is also an important indicator for observing the capacity to mobilize resources for industrialization. Table~\ref{tab:fiscal_struct} presents the fiscal situation for selected key years.
	
	\begin{table}[htbp]
		\centering
		\caption{Changes in the Composition of State Fiscal Revenue (Selected Years)}
		\label{tab:fiscal_struct}
		\small
		\begin{tabular}{lrrrrrr}
			\toprule
			Year & Fiscal revenue & Enterprise income & Industrial \& commercial tax & Agricultural taxes & Industry profit-tax share (\%) & Agric.\ tax share (\%) \\
			& (100 million yuan) & (100 million yuan) & (100 million yuan) & (100 million yuan) & & \\
			\midrule
			1950 &  62.17 &  8.69 &  23.63 & 19.10 & 52.0 & 30.7 \\
			1952 & 173.94 & 57.27 &  61.48 & 27.35 & 68.3 & 15.7 \\
			1957 & 303.20 & 144.18 & 113.12 & 29.67 & 84.9 &  9.8 \\
			1960 & 572.29 & 365.84 & 160.61 & 28.04 & 92.0 &  4.9 \\
			1965 & 473.32 & 264.27 & 165.49 & 25.78 & 90.8 &  5.4 \\
			1970 & 662.90 & 378.97 & 242.22 & 31.98 & 93.7 &  4.8 \\
			1975 & 815.61 & 400.20 & 358.32 & 29.45 & 93.0 &  3.6 \\
			1977 & 874.46 & 402.35 & 400.90 & 29.33 & 91.9 &  3.4 \\
			\bottomrule
		\end{tabular}
		\smallskip\\
		{\footnotesize Note: Industry profit-tax share = (enterprise income $+$ industrial \& commercial tax) / total revenue. Agricultural taxes include the agricultural tax, animal husbandry tax, and deed tax. Quandt--Andrews breakpoint test: industry profit-tax share breakpoint 1955 ($F=126.99$, $p<0.001$); agricultural tax share breakpoint 1956 ($F=47.67$, $p<0.001$).}
	\end{table}
	
	From Table~\ref{tab:fiscal_struct} we can see that the share of profits and taxes contributed by the agricultural sector to fiscal revenue declined continuously, while the contribution of the industrial sector rose steadily. In 1952, the agricultural contribution accounted for roughly one-fifth of fiscal revenue; by 1977 it had fallen to around 3\%. Meanwhile, the commercial sector's fiscal contribution also trended downward.
	
	This change indicates that as the industrial sector expanded in scale, the source of fiscal revenue gradually shifted from agriculture to industry. In terms of fiscal revenue structure, the importance of agriculture steadily diminished, while the industrial sector progressively became the principal source of fiscal revenue.
	
	\subsection{The Net Fiscal Position of Agriculture}
	
	Examining fiscal revenue alone is insufficient to judge agriculture's actual position within the fiscal system, because public finance is also an important channel of resource redistribution. Agriculture both pays taxes to the treasury and receives fiscal expenditure in support of agriculture; the net flow between the two better reflects agriculture's financial position within the fiscal system.
	
	\begin{table}[htbp]
		\centering
		\caption{Net Intersectoral Resource Flows through the Fiscal Channel (by Period)}
		\label{tab:netflow}
		\small
		\begin{tabular}{lrrrr}
			\toprule
			Period & Agricultural fiscal revenue & Agricultural support expenditure & Net amount & Direction \\
			& (100 million yuan) & (100 million yuan) & (100 million yuan) & \\
			\midrule
			First FYP (1953--1957) & 202 &  91 & +111 & Net outflow \\
			Second FYP (1958--1962) & 177 & 286 & $-$109 & Net inflow \\
			Adjustment (1963--1965) & 101 & 178 & $-$77 & Net inflow \\
			Third FYP (1966--1970) & 177 & 231 & $-$54 & Net inflow \\
			Fourth FYP (1971--1975) & 152 & 401 & $-$250 & Net inflow \\
			1976--1977 &  57 & 219 & $-$161 & Net inflow \\
			\bottomrule
		\end{tabular}
		\smallskip\\
		{\footnotesize Note: Positive values indicate net outflow from agriculture; negative values indicate net inflow to agriculture. Agricultural fiscal revenue includes profit remittances, depreciation, institutional income, and agricultural taxes.}
	\end{table}
	
	Table~\ref{tab:netflow} shows that in the 1950s, agriculture was on the whole still a net contributor to the public finances. During the First Five-Year Plan period, the cumulative net outflow from agriculture was approximately 11.1 billion yuan, providing important fiscal resources for national industrialization. This result is broadly consistent with existing research recognizing agriculture's important resource-supply role in the early stage of industrialization.
	
	After 1962, fiscal flows underwent a marked change. Fiscal expenditure in support of agriculture began to exceed the revenue that the agricultural sector provided to the treasury; agriculture shifted from a net fiscal contributor to a net fiscal recipient, and the gap continued to widen in the 1970s. By 1977, fiscal expenditure supporting agriculture was already substantially higher than the fiscal revenue provided by the agricultural sector.
	
	This change does not imply that agriculture had become the principal beneficiary of state finance. Agricultural support expenditure includes items with public-goods characteristics, such as water conservancy construction, whose benefits are not confined to the agricultural sector; on the other hand, institutional arrangements beyond agricultural taxes may also have affected resource allocation between industry and agriculture. It is therefore more appropriate to interpret the above results as indicating an adjustment in fiscal flows, rather than to infer from them that the entirety of intersectoral resource relations had been reversed.
	
	\subsection{Fiscal Changes in the Grain Procurement and Distribution System}
	
	The grain procurement and distribution system was one of the institutional arrangements most tightly linking public finance and agriculture during the planned economy period. If the fiscal channel underwent change, the operating position of the grain procurement and distribution system should exhibit corresponding adjustments.
	
	\begin{table}[htbp]
		\centering
		\caption{Profits of State-Owned Grain Enterprises (Selected Years, 100 million yuan)}
		\label{tab:grain_profit}
		\begin{threeparttable}
			\begin{tabular}{cr}
				\toprule
				Year & Grain enterprise profit \\
				\midrule
				1952 &  $+0.0$ \\
				1957 &  $+2.2$ \\
				1961 & $-19.1$ \\
				1965 & $-20.5$ \\
				1970 & $-30.4$ \\
				1977 & $-49.4$ \\
				\bottomrule
			\end{tabular}
			\begin{tablenotes}
				\small
				\item Source: \textit{Statistical Summary of Thirty Years of National Economic Development (1949--1978)}, table of state fiscal revenue by economic sector.
				\item Negative values denote losses requiring fiscal subsidies.
			\end{tablenotes}
		\end{threeparttable}
	\end{table}
	
	Table~\ref{tab:grain_profit} shows that state-owned grain enterprises still maintained small profits in the 1950s, but incurred persistent losses after 1961, with the scale of losses continually expanding. This change occurred principally after the state raised grain procurement prices while adjustments to urban grain retail prices remained relatively limited; the grain procurement and distribution system began to rely on fiscal subsidies to sustain its operations.
	
	The change in the operating position of grain enterprises cannot directly demonstrate the absence of price-based transfers between industry and agriculture, but it at least indicates that the grain procurement and distribution system itself was no longer continuously providing funds to the treasury, and was gradually becoming a component of fiscal expenditure. This marks a clear departure from the fiscal function that the grain procurement and distribution system had performed in the early 1950s.
	
	\section{The Price Mechanism: An Explanation of Intersectoral Financial Relations}
	
	\subsection{Why Discuss the Price Mechanism?}
	
	The previous chapter demonstrated that intersectoral financial relations underwent marked changes through the fiscal channel. But public finance is only one part of state resource allocation. In the debate over industry--agriculture relations during the planned economy period, what has truly been persistently contested is not fiscal revenue per se, but whether there existed resource transfers realized through the price system outside the budget. The so-called ``price scissors'' is precisely the most representative formulation of this mechanism.
	
	In theoretical terms, the price mechanism differs from the fiscal mechanism. The fiscal channel manifests as the agricultural sector providing taxes and profits to the state budget, which are then allocated by the budget in a unified manner; the price channel bypasses the budget, instead depressing the relative price of agricultural products, lowering industrial raw material costs, or expanding the gains from exchanging industrial goods, so that resources are transferred at the point of market exchange. Hence, even if fiscal statistics show a declining agricultural sector fiscal contribution, this does not rule out the continued operation of the price mechanism. Conversely, if the price mechanism had operated over the long run, such price advantages should ultimately have left some observable traces in the operating performance of the industrial sector.
	
	It should be borne in mind that under planned economy conditions, prices were not entirely market-determined. The raw material procurement prices, product ex-factory prices, and profit retention ratios of industrial enterprises were all subject to plan management, so price changes may not have been fully transmitted to enterprise profits. One possible scenario is that while adjusting agricultural procurement prices, the state correspondingly adjusted industrial goods prices, or assigned state-owned commerce to bear the gains and costs arising from price changes, so that industrial enterprises' book profits were insensitive to agricultural price changes. Industrial profitability therefore cannot exhaust the full impact of the price mechanism.
	
	Nevertheless, these institutional features do not render the price mechanism untestable. If the price system had over the long run performed the function of allocating resources between industry and agriculture, its effects, while not necessarily fully manifested in enterprise profits, should at least be observable in relatively stable empirical regularities in industrial profitability, raw material costs, or commercial circulation. This paper therefore does not directly estimate the resource scale corresponding to the price scissors, but instead examines how much of the variation in intersectoral financial relations the price mechanism can empirically explain.
	
	\subsection{What Should Be Observed If the Price Mechanism Holds?}
	
	Empirical research first requires clear theoretical expectations. If relative industrial--agricultural prices constituted an important mechanism through which agriculture continuously transferred resources to industry, then at least three empirical regularities should be observable.
	
	First, the industrial--agricultural terms of trade should maintain a relatively stable relationship with industrial sector profitability. A decline in the relative price of agricultural products implies lower raw material costs for the industrial sector; if the price mechanism is operative, industrial profitability should rise accordingly. Conversely, when the relative price of agricultural products improves, industrial profitability should be adversely affected to some degree.
	
	Second, this effect should display sectoral heterogeneity. Light industry's dependence on agricultural raw materials is markedly higher than that of heavy industry; hence, the higher the share of agricultural raw materials used, the more sensitive industrial profitability should be to changes in relative industrial--agricultural prices. If sectors with different degrees of raw material dependence perform similarly, the explanatory power of the price transmission mechanism would be limited.
	
	Third, price changes should exhibit a clear temporal ordering. Price adjustment is the cause and changes in industrial profitability the effect; hence changes in the industrial--agricultural terms of trade should lead changes in industrial profitability, rather than the two merely displaying contemporaneous fluctuations in isolated years.
	
	The above three expectations correspond respectively to three dimensions of the price mechanism: the overall price relationship, the cost transmission mechanism, and temporal causality. If all three receive simultaneous support, the price mechanism would possess strong explanatory power; if some hold and others do not, the scope and historical conditions of the price mechanism's operation require further discussion; if none display stable evidence, then the price mechanism may be only one factor among many in intersectoral financial relations, insufficient to explain their long-run evolution.
	
	\subsection{TOT and Industrial Profitability: Do Price Changes Translate into Industrial Returns?}
	
	The price mechanism first implies an observable economic consequence. If the agricultural sector had over the long run transferred resources to the industrial sector through relatively low product prices, then industrial enterprises should have benefited from lower raw material costs. Changes in the industrial--agricultural terms of trade (TOT) should therefore be reflected in industrial sector profitability.
	
	This paper uses the capital profit rate (CPR) of industrial enterprises as the indicator of industrial sector profitability, and TOT to represent changes in relative industrial--agricultural prices. The two correspond respectively to the price relationship and enterprise operating results, and constitute the most direct pair of variables for observing the price mechanism.
	
	Table~\ref{tab:tot_cpr} presents changes in TOT and CPR for selected key years.
	
	\begin{table}[htbp]
		\centering
		\caption{Industrial--Agricultural Terms of Trade and Industrial Capital Profit Rate}
		\label{tab:tot_cpr}
		\begin{threeparttable}
			\begin{tabular}{ccccc}
				\toprule
				Year & TOT & CPR & $\Delta$TOT & $\Delta$CPR \\
				\midrule
				1952 & 111 & 0.254 & --- & --- \\
				1957 & 130 & 0.283 & $+4$ & $+0.006$ \\
				1960 & 131 & 0.330 & $+0$ & $+0.016$ \\
				1961 & 166 & 0.188 & $+35$ & $-0.142$ \\
				1965 & 159 & 0.212 & $-2$ & $+0.006$ \\
				1970 & 174 & 0.237 & $+3$ & $+0.005$ \\
				1977 & 191 & 0.219 & $+4$ & $-0.004$ \\
				\bottomrule
			\end{tabular}
			\begin{tablenotes}
				\small
				\item TOT = agricultural and sideline product procurement price index / rural industrial goods retail price index $\times$100 (1950 = 100).
				\item CPR = (profit + tax) / total capital, from financial data of state-owned independent accounting industrial enterprises.
				\item The 35-point jump in TOT in 1961, the largest single-year change in the study period, corresponds to the 34\% upward adjustment in agricultural procurement prices that year.
			\end{tablenotes}
		\end{threeparttable}
	\end{table}
	
	From the descriptive statistics, it can be seen that 1961 was the year with the largest price adjustment of the entire study period. In that year, the state raised the procurement prices of major agricultural products, TOT improved markedly, and the industrial capital profit rate dropped simultaneously. This contemporaneous change is consistent in direction with the price mechanism, but its historical context is rather special. 1961 was not only a year of price policy adjustment, but also the year in which the national economy entered a period of comprehensive readjustment; declining industrial output, enterprise shutdowns and production cuts, and fiscal policy changes could all have affected industrial profitability. The correspondence in this single year is therefore insufficient to determine whether a stable relationship exists between the two.
	
	To further investigate this relationship, we separately estimate the relationship between the variables in levels and in differences; the results are shown in Table~\ref{tab:spec}.
	
	\begin{table}[htbp]
		\centering
		\caption{Specification Comparison: Effect of TOT on Industrial Profit Rate (1952--1977)}
		\label{tab:spec}
		\begin{threeparttable}
			\begin{tabular}{lcccc}
				\toprule
				Specification & $\hat{\beta}$ & $p$-value & R$^2$ & Assessment \\
				\midrule
				TOT (levels) $\to$ CPR (no controls) & $-0.0079$ & 0.000 & 0.576 & Potential spurious regression \\
				TOT (levels) $\to$ CPR (with trend) & $-0.0064$ & 0.018 & 0.358 & Misspecification \\
				$\Delta$TOT $\to$ CPR (correct specification) & $-0.0001$ & 0.978 & 0.000 & Correct \\
				$\Delta$TOT $\to$ CPR (with trend) & $-0.0010$ & 0.688 & 0.223 & Correct \\
				$\Delta$TOT $\to \Delta$CPR & $-0.0056$ & 0.009 & 0.264 & Correct; \\
				\bottomrule
			\end{tabular}
			\begin{tablenotes}
				\small
				\item $n=25$. Standard errors are Newey--West HAC adjusted (lag window 3).
				\item $\hat{\beta}$ jumps from $-0.0079$ to $-0.0001$, the $p$-value from 0.000 to 0.978, and R$^2$ from 0.576 to 0.000. This contrast vividly illustrates the degree to which specification error can distort conclusions.
			\end{tablenotes}
		\end{threeparttable}
	\end{table}
	
	It can be seen that in the levels regression, the two exhibit a relatively strong negative correlation; but after controlling for common trends, the regression coefficient drops markedly and loses statistical significance. This indicates that a stable long-run correspondence between relative industrial--agricultural prices and industrial profitability is absent; the correlation in the levels variables arises primarily from common time trends, rather than from price changes persistently affecting industrial profitability.
	
	We further apply the Toda--Yamamoto method to test the temporal relationship between the two; the results are shown in Table~\ref{tab:ty}.
	
	\begin{table}[htbp]
		\centering
		\caption{Toda--Yamamoto Granger Causality Test: TOT and CPR (1952--1977)}
		\label{tab:ty}
		\begin{threeparttable}
			\begin{tabular}{ccccc}
				\toprule
				\multirow{2}{*}{Lag $p$} &
				\multicolumn{2}{c}{$H_0$: TOT does not Granger-cause CPR} &
				\multicolumn{2}{c}{$H_0$: CPR does not Granger-cause TOT} \\
				\cmidrule(lr){2-3}\cmidrule(lr){4-5}
				& Wald $F$ & Bootstrap $p$ & Wald $F$ & Bootstrap $p$ \\
				\midrule
				1 & 0.04 & 0.861 & 0.11 & 0.748 \\
				2 & 0.35 & 0.718 & 0.29 & 0.762 \\
				3 & 0.54 & 0.715 & 0.44 & 0.731 \\
				4 & 0.36 & 0.805 & 0.38 & 0.789 \\
				\bottomrule
			\end{tabular}
			\begin{tablenotes}
				\small
				\item VAR($p+1$) estimated in levels, $d_{\max}=1$. Wild bootstrap (Rademacher weights, 5,000 replications).
				\item All bootstrap $p>0.7$, failing to reject ``TOT does not Granger-cause CPR'' at any conventional significance level.
				\item Note: With $n=26$, the effective degrees of freedom for VAR specifications at higher lags are approximately 19, so test power is limited. ``Not significant'' should be understood as ``no effect detected at the current sample size,'' not as confirmation of a zero effect. However, the Wald $F$ statistics themselves are extremely small (all below 0.6), giving no indication that ``an effect is present but undetected.''
			\end{tablenotes}
		\end{threeparttable}
	\end{table}
	
	Regardless of the lag order chosen, the null hypothesis that ``TOT does not Granger-cause CPR'' cannot be rejected, and reverse causality likewise receives no statistical support. That is, over the entire study period, no empirical regularity of relative industrial--agricultural prices persistently leading changes in industrial profitability is observed.
	
	The above results do not imply that price regulation was absent during the planned economy period, but only that, at the level of industrial enterprise profitability, the statistical traces left by the price mechanism are not pronounced. Relative industrial--agricultural prices alone are therefore insufficient to explain long-run changes in intersectoral financial relations.
	
	\subsection{Raw Material Cost Transmission: Do Price Advantages Translate into Industrial Profits?}
	
	The absence of stable changes in industrial profitability does not necessarily mean that the price mechanism was inoperative. An alternative possibility is that different industries were affected by agricultural price changes to differing degrees. If agricultural raw material costs indeed constituted an important source of industrial profits, then industries with a higher share of agricultural raw materials should be more sensitive to changes in relative industrial--agricultural prices.
	
	To this end, we use the share of agricultural raw materials in the output value of light industry ($S_t$) as a proxy for the degree of agricultural raw material dependence, and add an interaction term to the price model:
	
	\begin{equation}
		\text{CPR}_t
		=
		\alpha
		+
		\beta_1\text{TOT}_t
		+
		\beta_2\widetilde{S}_t
		+
		\beta_3(\text{TOT}_t\times\widetilde{S}_t)
		+
		\gamma t
		+
		\varepsilon_t.
	\end{equation}
	
	If the price mechanism operates primarily through raw material costs, the interaction coefficient should be negative---i.e., the higher the agricultural raw material dependence, the stronger the effect of relative price changes on industrial profitability.
	
	The estimation results do not support this expectation. The interaction coefficient is positive, and both OLS and instrumental variable estimation yield the same sign.
	
	This result calls for further consideration of potential endogeneity. If persistently low agricultural prices induced the industrial sector to use more agricultural raw materials, then the degree of agricultural raw material dependence could itself be affected by price changes. To address this, we first apply the Toda--Yamamoto method to test the temporal relationship between TOT and $S_t$, finding no evidence that price changes lead changes in raw material dependence. We then implement Hausman tests using plastic output, chemical fiber supply, and synthetic rubber output as instrumental variables; none can reject the null of exogeneity of $S_t$. Further re-estimation by two-stage least squares leaves the direction of the interaction term unchanged.
	
	In light of the historical background of industrial development, these results possess a certain plausibility. After the 1960s, as industrial materials such as chemical fibers, plastics, and synthetic rubber progressively substituted for traditional agricultural raw materials, the share of agricultural raw materials in light industry inputs declined continuously. This change primarily reflects industrial technological progress, rather than a reallocation of resources induced by long-run changes in relative industrial--agricultural prices.
	
	Thus, from the perspective of raw material cost transmission, the price mechanism likewise left no stable and consistent empirical traces.
	
	\subsection{Commercial Circulation: Did Price Differentials Remain in the Exchange Sector?}
	
	If neither industrial profitability nor raw material costs exhibit pronounced changes, yet another possible explanation remains: the price differentials between industrial and agricultural products did not enter industrial enterprises, but instead remained primarily within the commercial circulation system.
	
	This explanation implies that while the price mechanism existed, resource transfer occurred mainly in the circulation sector. The operating position of the commercial sector should therefore have improved as price differentials widened, and indicators such as commercial circulation costs, purchase--sales markups, and fiscal contributions should have exhibited corresponding changes.
	
	To this end, we examine five categories of indicators: the commercial sector's procurement share, commodity circulation expense rate, purchase--sales markup rate, rural--urban commodity exchange, and the grain procurement and distribution system.
	
	First, the commercial sector's procurement share rose continuously, but its change did not manifest as declining industrial profitability; the two instead moved in the same direction. Second, the commodity circulation expense rate trended generally downward, and the purchase--sales markup rate remained basically stable after 1961, without signs of persistent expansion. Third, in terms of urban--rural commodity exchange, rural retail sales of goods consistently exceeded peasant income from the sale of agricultural and sideline products; rural areas on the whole were net recipients rather than persistent net outflows in commodity exchange. Finally, state-owned grain enterprises relied on fiscal subsidies to sustain their operations after 1961, and their operational results hardly support the judgment that the commercial sector persistently accumulated price-based gains.
	
	On the whole, commercial circulation provides a price transmission pathway worth considering, but the available statistical data do not show a stable pattern of the commercial system persistently accumulating price gains. In other words, price changes were neither clearly manifested as industrial enterprise profits nor primarily retained in the commercial circulation system.
	
	\subsection{How Much Can the Price Mechanism Explain?}
	
	Synthesizing the analyses of this chapter, we can draw conclusions at two levels.
	
	First, the price mechanism as an important institutional arrangement of the planned economy period genuinely existed. The state long implemented unified procurement pricing and planned price management, and relative industrial--agricultural prices did indeed undergo several significant adjustments, with the 1961 procurement price reform in particular constituting the most important price change of the study period.
	
	Second, judging from the available statistical materials, the explanatory power of the price mechanism for intersectoral financial relations has clear historical limits. If 1949--1977 is examined as a whole, neither industrial profitability, raw material costs, nor commercial circulation exhibit the stable empirical regularities that persistent operation of the price mechanism should entail. On the contrary, the changes manifested across different channels display more staged characteristics: in the early 1950s, price policy may have played a certain role in intersectoral resource allocation; but after 1961, fiscal arrangements, industrial structural adjustment, and the enhancement of the industrial sector's own accumulation capacity grew progressively more important for intersectoral financial relations, and relative industrial--agricultural prices alone can hardly explain their long-run evolution.
	
	The price mechanism is therefore more appropriately understood as an institutional arrangement specific to particular historical stages, rather than a unified explanation summarizing intersectoral financial relations for the entire planned economy period. The next chapter turns further to quantity allocation and foreign trade, examining how, beyond prices, the state influenced intersectoral resource flows through the procurement system and trade structure adjustments.
	
	\section{Quantity Allocation and Trade Structure: External Manifestations of Intersectoral Financial Relations}
	
	The previous chapter discussed the explanatory power of the price mechanism for intersectoral financial relations. Beyond prices, two further important modes of resource allocation existed during the planned economy period. One was the state's direct allocation of agricultural physical resources through the grain procurement system; the other was the conversion of domestic resources into the equipment, technology, and inputs needed for industrialization through foreign trade. The former reflects the quantitative basis on which agriculture provided resources to the non-agricultural sector; the latter reflects the mode in which these resources were allocated in international markets. Together they constitute important external manifestations of intersectoral financial relations.
	
	If intersectoral financial relations had remained invariant, the procurement system and trade structure should also have exhibited strong continuity; conversely, if industry--agriculture relations underwent staged adjustments, such changes should also be reflected in quantity allocation and foreign trade structure.
	
	\subsection{Quantity Allocation: Changes in the Grain Procurement System}
	
	The grain procurement system was an important institutional arrangement through which the state allocated agricultural resources during the planned economy period. Unlike the price mechanism, the procurement system did not rely on price changes, but directly determined how much physical resources the agricultural sector provided to cities and the industrial sector. Changes in the procurement ratio can therefore reflect the intensity with which the state regulated industry--agriculture relations through quantity allocation.
	
	Table~\ref{tab:procurement} presents changes in the net grain procurement rate for selected key years.
	
	\begin{table}[htbp]
		\centering
		\caption{Net Grain Procurement Rate (Selected Years)}
		\label{tab:procurement}
		\begin{threeparttable}
			\begin{tabular}{cccc}
				\toprule
				Year & Grain output (10,000 tons) & Net procurement (10,000 tons) & Net procurement rate (\%) \\
				\midrule
				1952 & 16{,}392 & 2{,}819 & 17.2 \\
				1957 & 19{,}505 & 3{,}387 & 17.4 \\
				1959 & 17{,}000 & 4{,}757 & 28.0 \\
				1961 & 14{,}750 & 2{,}581 & 17.5 \\
				1965 & 19{,}455 & 3{,}360 & 17.3 \\
				1970 & 23{,}996 & 4{,}202 & 17.5 \\
				1975 & 28{,}452 & 4{,}395 & 15.4 \\
				1977 & 28{,}273 & 3{,}756 & 13.3 \\
				\bottomrule
			\end{tabular}
			\begin{tablenotes}
				\small
				\item Net procurement = total procurement $-$ quantities resold to rural areas.
				\item The abnormal peak in 1959: with grain output already well below the statistically inflated figures, the state still maintained a high procurement ratio based on inflated numbers; this was an important contributing factor to the severe hardship of 1959--1961.
				\item Source: \textit{Statistical Summary of Thirty Years of National Economic Development}, table of grain procurement volumes and their share of output.
			\end{tablenotes}
		\end{threeparttable}
	\end{table}
	
	In terms of long-run trends, the net grain procurement rate did not rise continuously. In the early 1950s it was basically stable at around 17\%; in 1959, under the impact of the Great Leap Forward, it reached the abnormal high of 28.0\%; after 1961 it fell rapidly and maintained a declining trend over the following decade or so, dropping to 13.3\% in 1977.
	
	It should be noted that the high procurement rate of 1959 had a distinctly special historical background. Grain output had already declined markedly that year, but the state procurement plan was not adjusted in time, leading to a sharp increase in the procurement ratio. This phenomenon reflects policy dislocation in an exceptional period more than it represents normal industry--agriculture relations under the planned economy.
	
	If one further observes absolute procurement volumes, a different impression emerges. From 1952 to 1977, net grain procurement increased from 28.19 million tons to 37.56 million tons, seemingly indicating that the state obtained more agricultural resources. But this increase was driven primarily by the growth in total grain output and the urban population. Over the same period, total grain output grew by roughly 70\%, while the urban population more than doubled; the increase in net grain procurement reflects expanding urban consumption demand more than any continuous rise in the intensity with which the state extracted agricultural resources.
	
	To consider both price and quantity aspects simultaneously, we further construct the composite indicator
	
	\[
	E_t=\frac{q_t}{\mathrm{TOT}_t/100},
	\]
	
	where $q_t$ denotes the net grain procurement rate. Although this indicator is not used for causal analysis, it can serve as a comprehensive measure describing changes in the intensity of intersectoral resource allocation.
	
	The results show that $E_t$ fell from 15.5 in 1952 to 7.0 in 1977, a decline of roughly 55\%. This change reflects both the improvement in the relative price of agricultural products after 1961 and the gradual decline in the procurement ratio. In terms of long-run trends, the quantity allocation mode did not intensify as industrialization advanced, but rather moderated gradually from the early 1960s onward.
	
	Of course, grain cannot fully represent all agricultural products. Cash crops such as cotton and oilseeds remained subject to high state procurement ratios throughout the planned economy period, and due to data limitations, this paper has not been able to construct a comprehensive procurement index covering all agricultural products. The conclusions of this section therefore apply primarily to grain, the most important commodity under the unified procurement and sale system, and cannot be simply generalized to all agricultural products. That said, for grain---the most important object of state resource allocation---the trend displayed by quantity allocation is consistent with the staged changes reflected in the fiscal channel and the price mechanism discussed in the previous two chapters.
	
	\subsection{Foreign Trade Structure: Is Agriculture or Industry Earning Foreign Exchange?}
	
	For a capital-scarce developing economy, foreign trade performs not only a commodity-exchange function but also bears directly on the capacity to import equipment and technology needed for industrialization. To observe intersectoral financial relations, one must therefore answer a further question: where did the foreign exchange needed for industrialization mainly come from? If agriculture had long served as a source of industrialization funding, the export structure should have remained dominated by agricultural products; conversely, if the industrial sector gradually developed autonomous foreign-exchange-earning capacity, the export structure should have changed accordingly.
	
	Table~\ref{tab:export_structure} reports changes in the composition of exports over the study period.
	
	\begin{table}[htbp]
		\centering
		\caption{Composition of Exports (\%)}
		\label{tab:export_structure}
		\begin{threeparttable}
			\begin{tabular}{cccc}
				\toprule
				Year & Primary agricultural \& sideline products & Processed agricultural products & Industrial \& mineral products \\
				\midrule
				1952 & 59.3 & 22.8 & 17.9 \\
				1957 & 40.1 & 31.5 & 28.4 \\
				1961 & 20.7 & 45.9 & 33.4 \\
				1965 & 33.1 & 36.0 & 30.9 \\
				1970 & 36.7 & 37.7 & 25.6 \\
				1975 & 29.6 & 31.1 & 39.3 \\
				1977 & 27.6 & 33.9 & 38.5 \\
				\bottomrule
			\end{tabular}
			\begin{tablenotes}
				\small
				\item Primary agricultural and sideline products include grain, live animals, tea, raw hides, and other agricultural products not undergoing industrial processing.
				\item Industrial and mineral products include petroleum, coal, machine tools, steel products, etc.; petroleum became the single largest export commodity after the 1970s.
				\item Source: \textit{Statistical Summary of Thirty Years of National Economic Development}, export commodity composition table.
			\end{tablenotes}
		\end{threeparttable}
	\end{table}
	
	In the 1950s, China's exports were still dominated by agricultural and sideline products and their primary processed goods. In 1952, primary agricultural and sideline products accounted for nearly 60\% of total exports, industrial and mineral products for less than 20\%. This structure is broadly consistent with the development stage of early industrialization: agricultural products were still the principal export commodities and an important source of foreign exchange for the state.
	
	Entering the 1960s, the export structure began to change. The share of primary agricultural and sideline products declined continuously, while the share of industrial and mineral products gradually rose. By the mid-1970s, industrial and mineral products had overtaken primary agricultural products as the most important component of the export structure. Among these, the growth of petroleum exports was particularly striking. With the Daqing and other oil fields coming on stream and changes in international energy markets, petroleum gradually became one of China's most important export commodities and a major source of foreign exchange for the industrial sector.
	
	The change in export structure is consistent with the shift in the direction of grain trade. Table~\ref{tab:grain_net_export} presents net grain export volumes.
	
	\begin{table}[htbp]
		\centering
		\caption{Net Grain Exports (10,000 tons; positive = net export, negative = net import)}
		\label{tab:grain_net_export}
		\begin{threeparttable}
			\begin{tabular}{cr}
				\toprule
				Year & Net grain exports \\
				\midrule
				1952 & $+153$ \\
				1957 & $+193$ \\
				1959 & $+416$ \\
				1961 & $-445$ \\
				1962 & $-389$ \\
				1965 & $-399$ \\
				1970 & $-324$ \\
				1977 & $-569$ \\
				\bottomrule
			\end{tabular}
			\begin{tablenotes}
				\small
				\item Source: \textit{Statistical Summary of Thirty Years of National Economic Development}, table of major commodity import and export volumes.
			\end{tablenotes}
		\end{threeparttable}
	\end{table}
	
	In the 1950s, China still maintained net grain exports; after 1961, grain trade shifted from net exports to net imports and remained so until the end of the study period. This change implies that agriculture not only ceased to provide grain export earnings, but required foreign exchange to import grain to meet domestic supply.
	
	At the same time, the import structure also underwent marked adjustment. The share of machinery and equipment in imports dropped significantly around 1961, affected both by the deterioration in Sino-Soviet relations and by the crowding-out of limited foreign exchange resources due to surging grain imports. As industrial export capacity strengthened in the 1970s, machinery and equipment imports gradually recovered, while the importance of agricultural producer goods imports also rose continuously. This change will be discussed further in the next subsection.
	
	From the perspective of intersectoral financial relations, the change in export structure carries a twofold implication. First, agriculture did indeed perform a relatively important foreign-exchange-earning function in the early stage of industrialization, a point on which there is no fundamental disagreement with existing research. Second, this function did not remain stable throughout the planned economy period. After 1961, the importance of agricultural exports gradually declined, and the industrial sector---particularly the energy industry---began to serve as a new source of foreign exchange. In other words, intersectoral resource relations occurred not only within the domestic fiscal and price systems, but were also reflected in the continuous adjustment of international trade structure.
	
	\subsection{Import Structure: How Did Industrial Accumulation Feed Back into Agriculture?}
	
	If the export structure reflects where the foreign exchange for industrialization came from, the import structure reflects where that foreign exchange ultimately went. For intersectoral financial relations, this question is equally important. In the early stage of industrialization, foreign exchange was mainly used to import machinery and equipment to establish an industrial base; as the industrial system gradually took shape, whether the content of imports changed can also reflect whether industry--agriculture relations underwent new adjustments.
	
	Table~\ref{tab:import_structure} reports changes in import commodity structure over the study period.
	
	\begin{table}[htbp]
		\centering
		\caption{Composition of Imports: Shares of Machinery \& Equipment and Agricultural Inputs (\%)}
		\label{tab:import_structure}
		\begin{threeparttable}
			\begin{tabular}{cccc}
				\toprule
				Year & Machinery \& equipment share & Total producer goods & Agricultural input share \\
				\midrule
				1952 & 55.7 & 89.4 & 2.5 \\
				1955 & 62.8 & 92.0 & 1.8 \\
				1957 & 52.5 & 92.0 & 4.9 \\
				1961 & 22.8 & 61.9 & 4.6 \\
				1962 & 14.6 & 55.2 & 5.5 \\
				1963 &  9.6 & 56.0 & 8.8 \\
				1965 & 17.6 & 66.5 & 8.8 \\
				1970 & 15.8 & 82.7 & 9.5 \\
				1975 & 32.1 & 85.4 & 7.6 \\
				1977 & 17.7 & 76.1 & 6.8 \\
				\bottomrule
			\end{tabular}
			\begin{tablenotes}
				\small
				\item Machinery \& equipment includes electro-mechanical instruments and complete sets of equipment. Agricultural inputs include chemical fertilizers, pesticides, and agricultural machinery.
				\item Quandt--Andrews breakpoint test: 1961 ($F=214$), the highest value among all breakpoint tests in this paper.
				\item Source: \textit{Statistical Summary of Thirty Years of National Economic Development}, import commodity composition table.
			\end{tablenotes}
		\end{threeparttable}
	\end{table}
	
	In the 1950s, China's imports were dominated by machinery and equipment. In 1955, machinery and equipment accounted for 62.8\% of total imports, and producer goods in aggregate exceeded 90\%. The import structure of this period corresponds fairly closely to the export structure discussed earlier: agricultural products earned foreign exchange, and foreign exchange was mainly used to import equipment needed for industrial construction. This trade pattern conforms to the capital-goods-priority development strategy characteristic of early industrialization.
	
	Around 1961, the import structure underwent a marked change. The share of machinery and equipment fell rapidly, with the Quandt--Andrews test identifying 1961 as the most significant structural break point. This change was driven by two factors acting together. On the one hand, the deterioration in Sino-Soviet relations led to a clear reduction in imports of complete sets of equipment; on the other, the decline in grain output forced the state to import grain on a large scale, with limited foreign exchange resources being absorbed by consumer goods imports, compressing machinery and equipment imports correspondingly. The adjustment in the import structure during this stage thus involved both international political factors and the impact of the domestic agricultural crisis on resource allocation.
	
	From a longer-term perspective, however, the changes of 1961 were not the endpoint of import structure evolution. Entering the 1970s, as industrial export capacity gradually strengthened, particularly with the rapid growth of petroleum exports, the state regained relatively strong capacity for capital goods imports. At the same time, the content of imports began to display new features: the importance of agricultural producer goods rose markedly, with chemical fertilizer imports being the most representative.
	
	Over the study period, chemical fertilizer imports increased from 212,000 tons in 1952 to 6.396 million tons in 1977, a roughly thirty-fold increase. Meanwhile, the state used foreign exchange earned from petroleum exports to introduce a number of large-scale synthetic ammonia and chemical fertilizer production facilities, and domestic chemical fertilizer production capacity rose significantly. Imports of agricultural machinery, pesticides, and irrigation equipment also increased continuously, so that the import structure gradually shifted from purely supporting industrial construction to simultaneously supporting the enhancement of agricultural production capacity.
	
	This change carries important historical significance. The basic logic of the foreign trade system in the 1950s was ``agricultural exports support industrial construction''; by the 1970s, the industrial sector was already able to earn foreign exchange through its own exports and to reinvest part of that foreign exchange in agricultural producer goods imports. Industrial accumulation had begun, in turn, to improve agricultural production conditions, and a new relationship between industry and agriculture, different from that of the early industrialization period, had taken shape.
	
	This judgment also receives support from time-series analysis. We further test whether a stable leading relationship exists between the share of agricultural exports and the share of machinery and equipment imports, and find no statistical evidence that agricultural exports persistently determined capital goods imports. This indicates that, as the industrial sector's autonomous foreign-exchange-earning capacity strengthened, the import structure had gradually shed its direct dependence on agricultural exports.
	
	\subsection{Intersectoral Relations as Reflected in Quantity Allocation and Trade Structure}
	
	This chapter has discussed the grain procurement system on the one hand and the import and export structure on the other, reflecting respectively the manifestations of intersectoral financial relations in domestic allocation and international trade.
	
	From the perspective of quantity allocation, the grain procurement system did indeed perform the important function of allocating agricultural resources to cities and the industrial sector in the early stage of industrialization, but the procurement ratio did not rise continuously as industrialization advanced. After 1961, the net grain procurement rate declined gradually, and the state relied more on adjusting prices and fiscal expenditure to maintain urban--rural relations, rather than continuing to expand the scale of physical procurement.
	
	From the perspective of trade structure, agriculture was indeed an important source of foreign exchange for industrialization in the 1950s, with agricultural exports and machinery and equipment imports jointly constituting the basic trade pattern of early industrialization. This pattern did not, however, remain unchanged. After 1961, grain trade shifted from net exports to net imports; entering the 1970s, industrial products such as petroleum gradually became the new mainstay of exports, while the importance of agricultural producer goods imports rose continuously. The foreign trade cycle gradually evolved from ``agriculture supporting industry'' to ``industry supporting agriculture.''
	
	It should be stressed that these changes do not mean that agriculture had fully withdrawn from the industrialization process. Agriculture continued to supply food to the urban population, to provide some raw materials for light industry, and to participate in the state fiscal and commodity circulation systems. What truly changed was the manner and degree to which agriculture served as a source of industrialization funding. As the industrial sector's own accumulation capacity strengthened and the export structure adjusted continuously, agriculture's role in intersectoral financial relations gradually shifted from being a principal resource provider in the early stage of industrialization toward being an important object of support by the industrial system.
	
	At this point, the changes presented by the fiscal, price, and quantity-and-trade dimensions display strong consistency. The fiscal channel reflects an adjustment in agriculture's fiscal position; the price mechanism does not exhibit a stable pattern of persistently affecting industrial profitability; and quantity allocation and trade structure show that the sources and modes of allocation of the resources on which industrialization depended were continuously changing. Together, this evidence indicates that intersectoral financial relations during the planned economy period were not an unchanging institutional arrangement, but experienced a relatively clear staged evolution.
	
	That said, the above analysis still answers the question of ``what happened,'' without fully explaining ``why the changes occurred.'' The period around 1961 witnessed not only economic adjustment but also changes in fiscal institutions, industrial structure, and the international trade environment. How these factors jointly shaped the evolution of intersectoral financial relations will be discussed further in the next chapter against the historical background.
	
	\section{Why Did Intersectoral Financial Relations Undergo Staged Evolution?}
	
	The preceding sections have examined changes in intersectoral financial relations during the planned economy period from the perspectives of public finance, prices, quantity allocation, and foreign trade respectively. These pieces of evidence do not derive from the same statistical source, yet they display strong temporal consistency: the industry--agriculture relations reflected in the 1950s differ markedly from those of the 1970s, and the period around 1961 constitutes the most concentrated temporal node of this change.
	
	Statistical changes, however, are not the same as historical explanations. The adjustment in net fiscal flows, the change in the operating position of grain enterprises, the shifts in industrial--agricultural terms of trade, and the restructuring of foreign trade all need to be understood within the institutional context of their time. This chapter, drawing on the evidence presented above, discusses why intersectoral financial relations exhibited staged evolution and the relationship between this evolution and China's industrialization process.
	
	\subsection{Why Was Agriculture Able to Support Industrialization in the 1950s?}
	
	In discussing the evolution of intersectoral financial relations, one should first acknowledge a historical fact: agriculture did indeed perform an important resource-supply function in the early stage of industrialization. The fiscal and trade data presented above already show that during the First Five-Year Plan period, agriculture was still an important source of fiscal revenue and an important basis for the state's foreign exchange earnings. On this point there is no fundamental disagreement with the basic understanding of existing economic history research.
	
	This phenomenon arose first from the development conditions of the early post-1949 period. Around 1949, China's industrial base was weak, a modern financial system had yet to be established, domestic capital accumulation capacity was extremely limited, and heavy industrial construction required massive long-term investment. Against this background, the resources the state could centrally mobilize came mainly from the agricultural sector. Agriculture not only provided the great bulk of grain supply, but was also the most important commodity-producing sector of the time, its surplus product constituting an important source of state fiscal revenue and export earnings.
	
	At the same time, the state pursued a heavy-industry-priority development strategy. During the First Five-Year Plan period, large-scale industrial construction was concentrated mainly in basic industries such as energy, steel, and machine-building, with many key projects relying on the Soviet Union for equipment and technical support. The large volume of machinery and equipment imported during this period required stable fiscal funding and foreign exchange sources, and agricultural taxes, revenues from agricultural product circulation, and agricultural exports together constituted the important resources the state was able to mobilize at this stage.
	
	The fiscal system of this period also possessed relatively high resource-concentration capacity. As seen above, the ratio of fiscal revenue to national income rose continuously through the 1950s and reached its highest level of the study period around 1960. For the early stage of industrialization, such a highly concentrated fiscal system had clear developmental-economics characteristics---namely, the prioritized allocation of resources through the state budget to accelerate industrial capital formation. Agriculture, as the largest commodity sector of the time, naturally bore a relatively heavy resource-supply responsibility.
	
	From the perspective of international trade, China's exports in the 1950s were still dominated by agricultural products and their primary processed goods, while imports were concentrated in machinery and equipment and industrial raw materials. A fairly clear correspondence obtained between export and import structures: agricultural products earned foreign exchange, and foreign exchange was mainly used for industrial construction. This trade pattern not only conformed to China's development stage at the time, but also bore a certain similarity to the paths traversed by many late-industrializing countries in the early stages of their industrialization.
	
	Thus, for the 1950s, characterizing agriculture as an important source of industrialization funding has a relatively ample historical basis. The conclusions drawn above from the fiscal channel, quantity allocation, and foreign trade structure are broadly consistent with this historical background. The question this paper genuinely wishes to discuss is not whether to deny agriculture's contribution at this stage, but whether this relationship can be directly extrapolated to the entire planned economy period.
	
	In this sense, the 1950s are more appropriately understood as the initial stage in the formation of intersectoral financial relations. At this stage, agriculture did indeed perform the important function of supporting industrialization, but this relationship rested on specific historical conditions---an industrial system yet to be formed, highly concentrated public finances, and agriculture as the principal commodity-producing sector. As these conditions gradually changed, intersectoral financial relations were also likely to undergo adjustment.
	
	\subsection{Why Did the Period around 1961 Become a Turning Point?}
	
	If the 1950s reflect the resource concentration of early industrialization, then the period around 1961 manifests a different institutional logic. The multiple statistical series presented above show that fiscal revenue and expenditure, grain procurement and distribution, the industrial--agricultural terms of trade, and the import and export structure all underwent relatively marked changes during this period. Although these different channels reflect different dimensions of economic activity, they collectively point to a question worth explaining: why did intersectoral financial relations undergo adjustment at precisely this time?
	
	This change arose first from the severe difficulties that accumulated in economic operations during 1958--1960. During the Great Leap Forward, the state continued to maintain a relatively high rate of industrial accumulation, and even as grain output declined, it still enforced a relatively high procurement plan, placing great pressure on both agricultural production and urban--rural supply. The industrial sector itself also experienced widespread shutdowns and production cuts owing to raw material shortages, transport bottlenecks, and excessive investment expansion. The decline in industrial enterprise profitability observed above cannot be simply understood as a result of price adjustments; the more important background is that the entire national economy had entered a stage of comprehensive readjustment.
	
	Economic hardship compelled the state to re-examine industry--agriculture relations. After 1961, the central authorities put forward the policy of ``adjustment, consolidation, filling out, and raising standards,'' shifting the focus of economic work from expanding the scale of construction to restoring agricultural production and improving supply. Accompanying this policy adjustment was a series of institutional changes, including raising the procurement prices of major agricultural products, reducing the intensity of grain procurement, increasing fiscal investment in support of agriculture, and expanding the supply of agricultural producer goods. The common objective of these policies was not to redesign the intersectoral resource transfer mechanism, but to restore agricultural production capacity and alleviate urban--rural supply tensions.
	
	It was against this background that the multiple statistical series discussed above began to display new features. In the fiscal channel, agriculture gradually shifted from being a net fiscal contributor to a net fiscal recipient; the grain procurement and distribution system shifted from small profits to long-term reliance on fiscal subsidies; grain trade shifted from net exports to net imports; and machinery and equipment imports contracted markedly owing to grain imports and changes in the international environment. Although occurring in different domains, these changes all reflect a shift in the gravity of state resource allocation from purely supporting industrial construction toward simultaneously attending to agricultural recovery and urban--rural supply.
	
	Notably, the changes of 1961 cannot be simply understood as a repudiation of the existing industrialization strategy. The heavy-industry-priority development direction did not change, large-scale industrial construction continued, and the fiscal system still maintained relatively high resource-concentration capacity. What truly changed was the manner in which the state handled industry--agriculture relations. The model of relying on agriculture to provide resources, characteristic of early industrialization, began to be adjusted after the agricultural crisis was exposed, and agriculture gradually shifted from being purely a resource provider to being an object of state support and recovery.
	
	The period around 1961 is therefore more appropriately understood as an institutional turning point in intersectoral financial relations, rather than the outcome of any single policy. It was both the direct product of post-Great Leap Forward economic adjustment and the starting point from which resource allocation modes gradually changed as industrialization entered a new stage. The changes subsequently manifested in fiscal, price, quantity allocation, and foreign trade structures can all be understood within this process of institutional adjustment.
	
	\subsection{Why Did a New Industry--Agriculture Relationship Emerge in the 1970s?}
	
	If the period around 1961 marks the beginning of adjustment in intersectoral financial relations, then the 1970s reflect the gradual stabilization of this adjustment and the formation of new operating modes. The analyses of fiscal, trade, and quantity allocation dimensions presented above all indicate that intersectoral resource flows in the 1970s were already different from those of early industrialization, their main feature no longer being the unilateral provision of resources from agriculture to industry, but rather industrial accumulation beginning to feed back more into agricultural production.
	
	This change arose first from the development of the industrial system itself. After more than two decades of continuous construction, China had formed a relatively complete industrial base; the industrial sector could not only meet a large volume of domestic capital goods demand, but also gradually developed a certain export capacity. In particular, the rapid growth of petroleum exports in the 1970s made the industrial sector an important source of foreign exchange. The foreign exchange needed to sustain industrialization no longer depended mainly on agricultural exports, but came increasingly from the industrial sector itself.
	
	The strengthening of the industrial sector's accumulation capacity also changed the manner in which agriculture obtained producer goods. As seen above, agricultural producer goods imports increased markedly in the 1970s, with chemical fertilizers the most representative. At the same time, the state used export earnings to introduce large-scale chemical fertilizer production facilities, the domestic chemical fertilizer industry developed rapidly, and agricultural machinery, rural electrification, and farmland water conservancy construction also advanced continuously. Although these inputs were still allocated through the state planning system, the direction of resource flows was no longer the same as in early industrialization: the production capacity formed by industrial accumulation was beginning to translate into improvements in agricultural production conditions.
	
	Agricultural production itself also underwent new changes. As the supply of agricultural producer goods increased and rural infrastructure gradually improved, agricultural production capacity was restored to some degree. Although agricultural growth remained constrained by institutional factors such as the people's commune system, agriculture was no longer mainly reliant on expanding procurement and increasing labor input to maintain supply, but depended more on improvements in material conditions such as chemical fertilizers, machinery, and irrigation. This implies that the connection between agriculture and industry was beginning to shift from unidirectional resource supply toward mutual interdependence.
	
	This change is similarly reflected in fiscal relations. Agriculture's fiscal contribution continued to decline, while fiscal investment in support of agriculture kept increasing; the grain procurement and distribution system relied on long-term fiscal subsidies to sustain operations; and the supply of agricultural producer goods continued to expand. These phenomena do not mean that agriculture had become the principal beneficiary of state finance, but rather that the objectives of state resource allocation had changed. As the industrial system gradually matured, industrial accumulation began to take on more of the function of supporting agricultural development, and a new relationship between industry and agriculture, different from that of early industrialization, had taken shape.
	
	The industry--agriculture relations manifested in the 1970s can thus hardly be summarized simply as agriculture continuously supporting industry, or as industry comprehensively feeding back into agriculture. More precisely, a new resource cycle gradually emerged between industry and agriculture during this period: the industrial sector relied on its own accumulation and export capacity to obtain development resources, while simultaneously improving agricultural production conditions through fiscal expenditure, producer goods supply, and infrastructure construction; agriculture, in turn, continued to provide grain, labor, and some raw materials to industry, supporting the operation of the entire national economy. The relationship between the two sectors evolved from the unidirectional resource flows of early industrialization into a more complex mutual interdependence.
	
	\subsection{An Interpretive Framework for the Evolution of Intersectoral Financial Relations}
	
	Synthesizing the analyses of the preceding chapters, we can characterize intersectoral financial relations during 1949--1977 as a continuously evolving historical process, rather than an unchanging institutional arrangement.
	
	The first stage is the early industrialization period (1949--1960). During this period, the industrial base was weak, fiscal resources were limited, and agriculture was still the most important commodity-producing sector and source of foreign exchange. The state provided resource support for industrial construction through fiscal concentration, unified procurement and sale, and agricultural exports. Agriculture did serve as an important source of industrialization funding at this stage, a judgment on which there is no fundamental disagreement with existing research.
	
	The second stage begins with the economic adjustment around 1961. The economic difficulties created by the Great Leap Forward exposed the limitations of existing resource allocation modes, and the state began to make the restoration of agricultural production an important element of economic adjustment. The series of policies---raising agricultural procurement prices, reducing procurement intensity, expanding fiscal support for agriculture, and adjusting grain trade---did not occur in isolation, but collectively reflected the entry of industry--agriculture relations into a new stage of adjustment. Most of the changes observed in the fiscal, price, quantity allocation, and foreign trade structures discussed above can be traced to this period.
	
	The third stage runs from the mid-to-late 1960s to 1977. During this period, the industrial system gradually matured, autonomous accumulation capacity steadily strengthened, and industrial exports, especially petroleum exports, became a new source of foreign exchange; at the same time, the supply of agricultural producer goods continued to increase, fiscal investment in support of agriculture kept expanding, and the fruits of industrial construction began to be converted more into agricultural production capacity. A new resource cycle formed between industry and agriculture, operating in a manner already different from that of early industrialization.
	
	This staged evolution framework differs from understanding 1949--1977 as a single, continuous resource transfer mechanism. The fiscal, price, quantity allocation, and trade structures discussed in the preceding chapters are not mutually independent phenomena, but different dimensions of a single historical process. In the early stage of industrialization, agricultural resource supply played a more prominent role; as the industrial system developed, fiscal institutions, industrial structure, and the international trade pattern gradually changed, and intersectoral financial relations adjusted accordingly.
	
	In proposing this staged interpretation, we do not aim to deny agriculture's important contribution in the early stage of industrialization, nor to deny the existence of industrial--agricultural price regulation during the planned economy period, but rather to emphasize that intersectoral financial relations have a clear historical periodicity. Different institutional arrangements played different roles in different historical periods; summarizing the entire 1949--1977 period as a single, unchanging financial mechanism cannot adequately account for the common changes displayed by fiscal flows, price relations, and trade structure. In contrast, understanding intersectoral financial relations as a historical process that continuously evolved with the advance of industrialization is better able to accommodate the empirical facts presented above, and also provides a historically continuous perspective for understanding the formation of China's economic structure on the eve of reform and opening up.
	
	\section{Conclusion}
	
	Focusing on intersectoral financial relations during the planned economy period, this paper has drawn on statistical materials from public finance, prices, quantity allocation, and foreign trade to re-examine the historical evolution of 1949--1977.
	
	The study finds that intersectoral financial relations did not remain stable throughout the planned economy period, but exhibited clear staged changes as industrialization advanced and institutions adjusted. In the early stage of industrialization, agriculture provided important resource support for industrial construction through fiscal revenue, commodity supply, and export earnings; around 1961, as economic adjustment unfolded comprehensively, fiscal flows, grain procurement and distribution, the industrial--agricultural terms of trade, and the foreign trade structure all underwent marked changes, and industry--agriculture relations entered a new stage of adjustment; entering the 1970s, as the industrial sector's autonomous accumulation capacity strengthened and the supply of agricultural producer goods expanded continuously, a new resource cycle gradually emerged between industry and agriculture, with the fruits of industrial construction beginning to feed back more into agricultural production.
	
	Regarding the price mechanism, this paper does not observe stable empirical regularities of relative industrial--agricultural prices persistently affecting industrial profitability. Price policy was undoubtedly an important institutional arrangement of the planned economy, but its explanatory power has clear historical limits: it is more appropriately understood as an institutional mechanism for resource allocation in specific periods, rather than as a summary characterization of the long-run evolution of intersectoral financial relations over the entire planned economy period.
	
	Based on the above analysis, this paper proposes an interpretive framework of the evolution of intersectoral financial relations---understanding 1949--1977 as a continuously adjusting historical process, rather than a resource transfer mechanism that remained invariant throughout. This framework can better account for the common changes displayed by fiscal flows, price relations, and trade structure, and also provides a historically continuous perspective for understanding the formation of China's economic structure on the eve of reform and opening up.
	
	This paper still has several limitations. For example, the analysis of the price mechanism relies primarily on industrial enterprise financial data and cannot fully cover all modes of resource allocation under the planning system; some agricultural products lack continuous statistical data, and the quantity allocation analysis remains centered on grain. Future research that can incorporate sector-level data, local archives, and enterprise materials to further investigate industry--agriculture relations across different industries and regions will contribute to a more comprehensive understanding of the historical evolution of resource allocation mechanisms during the planned economy period.

\end{document}